\documentclass[aps,pra,amssymb,showpacs]{revtex4}

\def\be{\begin{equation}}
\def\ee{\end{equation}}
\def\beq{\begin{eqnarray}}
\def\eeq{\end{eqnarray}}

\usepackage{psfig}
\begin{document}

\title{Green functions for generalized point interactions in $1D$:
A scattering approach}

\author{Alexandre G. M. Schmidt}
\email{schmidt@fisica.ufpr.br}
\author{Bin Kang Cheng}
\author{M. G. E. da Luz}
\email{luz@fisica.ufpr.br}
\affiliation{Departamento de F\'{\i}sica, Universidade Federal do
Paran\'a, C.P. 19044, 81531-990 Curitiba-PR, Brazil}

\date{\today}

\begin{abstract}
Recently, general point interactions in one dimension has been
used to model a large number of different phenomena in quantum
mechanics. Such potentials, however, requires some sort of
regularization to lead to meaningful results. The usual ways to do
so rely on technicalities which may hide important physical
aspects of the problem. In this work we present a new method to
calculate the exact Green functions for general point interactions
in 1D. Our approach differs from previous ones because it is based
only on physical quantities, namely, the scattering coefficients,
$R$ and $T$, to construct $G$. Renormalization or particular
mathematical prescriptions are not invoked. The simple formulation
of the method makes it easy to extend to more general contexts,
such as for lattices of $N$  general point interactions;
on a line; on a half-line; under periodic boundary conditions;
and confined in a box.

\end{abstract}


\maketitle

\section{Introduction}

Point interactions in one or more dimensions have been of great interest
in quantum physics and one can regard their relevance as being three-fold.
First, to model different phenomena such as:
energy band structure in ordered lattices \cite{kp};
emergence of quantum chaos \cite{qchaos};
Aharonov-Bohm effect in spin-1/2 particles \cite{ab,park1},
duality between fermionic and bosonic systems \cite{fb-duality}, etc.
Second, to allow exact closed analytical solutions in quantum mechanics
\cite{exact}, which are usually rare, however quite useful. For
instance, one of the few exactly solved many-body quantum problem
is 1D identical particles interacting by pairwise
$\delta$-function potentials \cite{bethea}. More recently,
progress in extending such solutions to general point interactions
has also been reported \cite{pibethea,lauro}.

The third relevance of such potentials is to provide relative
simple situations where development of regularization procedures
are in order. This is important not only due the applications
mentioned above, but also because such techniques may be extended
to more complex and general contexts  \cite{general}, e.g.,
anyons statistics, vortices and topological structures
in scattering, quantum field theory, etc.

We can state the non-relativistic quantum problem of a point
interaction as the following.
Consider a $d$-dimensional Hamiltonian,
written formally as $H({\bf r}) = H_0({\bf r}) + \Xi({\bf r}; {\bf
r}_0)$. $H_0 = -\nabla_{\bf r}^2 + V({\bf r})$ is a well defined
self-adjoint ``unperturbed'' Hamiltonian and $\Xi$ represents a
general point interaction potential located at ${\bf r}_0$. One
may have interest in the wave function, the Green function or
its Fourier transform, the propagator, which satisfy to $H({\bf
r}) \psi({\bf r}) = E \psi({\bf r})$, $(E - H({\bf r}_f)) G({\bf
r}_f,{\bf r}_i;E) = \delta({\bf r}_f - {\bf r}_i)$, and $K({\bf
r}_f,{\bf r}_i;t) = (2 \pi i)^{-1} \int dE \exp[- i E t] G({\bf
r}_f,{\bf r}_i;E)$, respectively.
The whole issue is to ask if $H$ is a
self-adjoint operator and the above equations lead to physical
meaningful $\psi$, $G$ and $K$, for instance, the quantum state
has an unique time evolution.

If the point interaction is the usual delta-function,
$\gamma \delta({\bf r} - {\bf r}_0)$, the
answer is negative for $d$ equal to 2 and 3. In 1D, this is
also the case for more singular point interactions such as
delta prime-function, $\gamma \delta'(x-x_0)$.
Different approaches are then used to regularize $H$.
Methods such as formal self-adjoint extensions (see, for instance,
Refs. \cite{zorbas,exact,park1}),
series expansion \cite{grosche1} and renormalization
\cite{renormalization} have been used in two or more dimensions.
For the one-dimensional case the situation is far more rich.
This is so because in 1D there is a four parameter family solution
\cite{exact, fourfamily} for the problem (see next Section),
so different types of discontinuities for point interactions are
possible.
In studying such family solution,
functional methods \cite{functional}, non-relativistic limit of the
Dirac's equation \cite{dirac} and self-adjoint extension \cite{park2}
have been used to calculate Green functions and propagators.

From the mathematical point of view, all the mentioned procedures
are quite ingenious. However, they rely on technicalities which
may hide important aspects of the problem, making hard to
understand the physical meaning of relevant quantities, like
potential strengths and scattering amplitudes. In fact, in some
cases the regularization of $H$ leads to a renormalization of
some parameters related to $\Xi$, which become then dependent on
the energy and the spatial position \cite{park2}.
So, one may find difficult to interpret
scattering by point interactions in terms of the so called inverse
scattering problem \cite{inverse}. Furthermore, some of the above
methods are cumbersome to apply to more general cases, e.g., for
many point interactions of different types, or for certain $V$'s.

Hence, it would be desirable to have a more pedestrian treatment
for the regularization of single point interactions, as well as to
calculate quantum objects, such as Green functions, directly from
concrete physical quantities, instead of renormalized ``bare''
parameters (difficult to identify in a real system). Moreover, the
method should be simple enough to be extended to more general
situations. Actually, such a point of view has already been used
to discuss $\delta$ interactions in two dimensions \cite{weaver}.
In the present work we show how to do that for $N$ general point
interactions in 1D under different boundary conditions.

Our paper is organized as the following. In Section II we show how
to characterize any general point interaction through its
scattering amplitudes. Then, with the help of some results known
in the literature we are able to readily write down the exact
Green function for the problem. In Section III, from a multiple
scattering approach, we extend the calculations and obtain the
exact $G$ for an array of $N$ general point interactions under
different conditions, namely, on a line, on a half-line, confined
in a box (with different boundary conditions at the walls) and
finally for periodic boundary conditions. Possible physical
applications for all these systems are briefly discussed. Finally,
in Section IV we drawn our final remarks and conclusion.

\section{The Scattering amplitudes characterization of a general
point interaction and the Green function}

It is a well-known fact that solving one-dimensional Schr\"odinger
equation for a delta potential located at the origin, $\delta(x)$,
is equivalent to the boundary conditions
($\psi'(x)\equiv d\psi/dx$)
\be \label{condicao-geral}
\left(\matrix{\psi({0}^+) \cr
\psi'({0}^+)}\right) =
\omega\left( \matrix{ a & b \cr
c & d  }\right) \left(\matrix{\psi({0}^-)\cr
\psi'({0}^-)}\right), \ee
where the parameters values are $a = d = \omega = 1$, $b = 0$ and
$c=\gamma$, with $\gamma$ the potential's strength.
This boundary condition can be obtained from the Schr\"odinger equation
by imposing that the wave function is continuous at $x=0$.
However, the same does not apply if the potential in question is the
delta-prime, $\delta'(x)$: the boundary condition satisfied by
$\psi(x)$ cannot be determined from the Schr\"odinger equation.
The only condition known {\it a priori} is the one $\psi'(x)$
fulfills, namely, $\psi'(x)$ is continuous at $x=0$.
For this very reason self-adjoint extension is
invoked \cite{exact,bonneau}. So, one can solve an equivalent
problem to Schr\"odinger equation with delta-prime potential
imposing (\ref{condicao-geral}) with
$c = 0$, $a = d = \omega = 1$ and $b=\gamma$.

The above two examples do not represent all possible one-dimensional
point interactions. In fact, through self-adjoint extension technique
it is shown that the most general case is the one in which
\be \label{restrictions}
|\omega|=1 \qquad \mbox{and} \qquad a d - b c = 1, \ \mbox{with} \ \
a, b, c, d \ \ {\mbox{all real}}.
\ee

An important aspect of this prescription is that it does not allow
a concrete realization for generalized point interactions.
In other words, it does not lead to a unique function depending on
$(a,b,c,d,\omega)$ which reproduces the whole boundary conditions
given in (\ref{condicao-geral}) and (\ref{restrictions}).
So, we cannot write a Hamiltonian $H=H_0+\Xi(x)$, since one does not
know a single form for the potential $\Xi(x)$
(actually, a different procedure is to represent a generalized point
interaction by making compositions of triple delta functions and then
taking certain limits \cite{shigehara-pla}, which, however,
also cannot be put in the form of an usual potential).

An alternative way to characterize the boundary conditions
(\ref{condicao-geral}) is through the scattering amplitudes. Let a
plane wave function, of wave number $k$ and incident either from
the left $(+)$ or right $(-)$, be written as \be
\label{scattering} \psi^{(\pm)}(x) = \frac{1}{\sqrt{2 \pi}} \times
\left\{\matrix{\exp{[\pm ikx]} + {R}^{(\pm)} \exp{[\mp ikx]}, & x
{< \atop >} 0 \cr & \cr {T}^{(\pm)} \exp{[\pm ikx]}, & x {> \atop
<} 0} \right. . \ee $\psi$ satisfies to $-d^2 \psi(x)/dx^2 = k^2
\psi(x)$ for $x\neq 0$. Now, if we choose for the scattering
amplitudes ($\theta^{(+)} = a d - b c$ and $\theta^{(-)} = 1$) \be
{R}^{(\pm)} = \frac{c \pm ik(d-a) + bk^2} {-c + i k(d+a) + b k^2},
\qquad {T}^{(\pm)} = \frac{2 i k \omega^{\pm 1} \theta^{(\pm)}}
{-c + i k(d+a) + b k^2 }, \label{RT-zero} \label{rt} \ee we find
that (\ref{scattering}) satisfies to the boundary conditions
(\ref{condicao-geral}). Furthermore, by imposing \cite{inverse}
\be \label{rtrel} |{R}|^2+|{T}^{(\pm)}|^2=1,\qquad {{R}^{(+)}}^*
{T}^{(+)} + {{T}^{(-)}}^* {R}^{(-)} = 0, \qquad {{R}_k^{(\pm)}}^*
= {R}^{(\pm)}_{-k}, \qquad {{T}_k^{(\pm)}}^* = {T}^{(\mp)}_{-k},
\ee the parameters must necessarily obey to (\ref{restrictions}).
The conditions in (\ref{rtrel}) assure important properties, e.g.,
probability conservation and the existence of the scattering
inverse problem (see \cite{inverse} for a detailed discussion).
Furthermore, if one also requires time-reverse invariance, which
is translated into the relation ${T}^{(+)} = {T}^{(-)}$ (what we
are not imposing in this work), then one should choose $\omega=\pm
1$.

From all these results we see that there is a complete equivalence
in defining the most general point interaction through
(\ref{condicao-geral})-(\ref{restrictions}) or by specifying its
scattering amplitudes (\ref{rt})-(\ref{rtrel}). We also observe
that eventually we may have bounded states for a given point
interaction potential depending on its parameters. In such case
the quantum amplitudes $R$ and $T$ have poles at the upper-half of
the complex plane $k$, corresponding to the eigenvalues. The
eigenfunctions can then be obtained from an appropriate extension
of the scattering states to those $k$'s values \cite{schiff}.

The exact Green function for arbitrary potentials of compact
support have been obtained in \cite{marcos1}, with an extension
for more general potentials presented in \cite{marcos2}. For the
derivations in \cite{marcos1}, it is necessary that the $R$ and
$T$ satisfy certain conditions, which indeed are the ones in
(\ref{rtrel}). Thus, based on \cite{marcos1} we can calculate the
Green function for general point interactions by using the
reflection and transmission coefficients, which for their very
construction carry information on boundary conditions and are
relevant quantities with a clear physical interpretation (being
the measured quantities in real situations \cite{stockmann}).

So, from \cite{marcos1} we can readily write down the exact Green
function as the following. By defining $G_{+-}$ for $x_f > 0 >
x_i$, $G_{-+}$ for $x_i > 0 > x_f$, $G_{++}$ for $x_f, \ x_i > 0$
and $G_{--}$ for $x_f, \ x_i < 0$, we have \be
\label{green-aberto} G_{\pm \mp} = \frac{1}{2ik} T^{(\pm)} \exp[ik
|x_f - x_i|], \qquad\qquad G_{\pm \pm} = \frac{1}{2ik} \left[
\exp[ik |x_f - x_i|] + R^{(\pm)} \exp[i k( |x_f| +|x_i|)]\right] .
\ee

The Green function for $\delta(x)$ and $\delta'(x)$ were
calculated, respectively, in \cite{blinder} and
\cite{grosche1,dirac,park2}. As we show in the Appendix A, the $G$'s in
(\ref{green-aberto}) do reduce to those cases if we assume for
the parameters the appropriate values as previously discussed.

\section{Arbitrary number of potentials and different
boundary conditions}

The advantage of the present method is that it can be easily used
to calculate the exact Green function for arbitrary (finite)
number of different point interactions, both on the infinite line
or for periodic boundary conditions. Also, we can obtain the $G$'s
for restricted systems such as $N$ potentials on a half-line or
confined inside an infinite box, with different boundary
conditions at the border walls.

There are many reasons to study such problems.
For instance, for $N$ disordered general point interactions on a
line one may have interest in the propagation of wave packets through
the lattice in order to analyze the influence of more singular
potentials in the usual scale theory of localization
\cite{localization}.
For periodic boundary conditions, we recall recent experiments
using a waveguide filled with localized scatterers in a circular
setup \cite{stockmann2}. The dynamics of the microwaves is
analogous to our 1D quantum system. They measure the
transmission amplitudes along the waveguide
and observe the Hofstadter butterfly \cite{hofstadter},
which has a fractal structure (a Cantor set).
Similar systems are used to study the quantum dynamics of eigenvalues
on the change of some external parameter \cite{stockmann3}.
Also, periodic boundary conditions were used \cite{fb-duality} to study
duality properties of point interactions in systems of bosons and
fermions.

\subsection{Green function for arbitrary number of point
interactions on the line}

Consider $N$ point interaction potentials, each characterized
by its location $y_n$ (with $y_{n-1} < y_n$) and the set of parameters
$\{a_n,b_n,c_n,d_n,\omega_n\}$. The quantum coefficients are
$R^{(\pm)}_n(y_n)$ and $T^{(\pm)}_n(y_n)$. For a potential located
at $y_n$, it is easy to see that $R_n^{(\pm)}(y_n) =
R_n^{(\pm)} \exp{(\pm 2ik y_n)}$ and $T_n^{(\pm)}(y_n) = T_n^{(\pm)}$,
where $R_n^{(\pm)}$ and $T_n^{(\pm)}$ are given by (\ref{RT-zero}).
In other words, the reflection coefficients change by a phase factor
while transmission coefficients remain the same.

Now, we can obtain the exact $G_{j m}(x_f,x_i;k)$ for $y_{m} < x_i
<  y_{m+1}$ and $y_{j} < x_f < y_{j+1}$, (for arbitrary $m$ and
$j$) by using the multiple scattering approach introduced in
\cite{marcos1,marcos2}. We only discuss the case of $x_i < x_f$,
see Fig. 1. The idea is simple, $G$ is given by $(2 i k)^{-1}
\sum_{\mbox{\scriptsize s.p.}} W_{\mbox{\scriptsize s.p.}} \exp[i
S{\mbox{\scriptsize s.p.}}(x_f,x_i;k)]$. The sum is performed over
all possible scattering paths (s.p.) joining the end points, with
$S_{\mbox{\scriptsize s.p.}}$ and $W_{\mbox{\scriptsize s.p.}}$
their actions and the amplitudes (or weights). For each scattering
path, $S$ is calculated from the free propagations between the
potentials. $W$ is the product of the quantum coefficients (up to
the phase factors) gained each time the particle is scattered off
by a point interaction. When the particle hits the potential $n$,
it can be reflected (getting an amplitude factor $R_n$) or
transmitted (getting an amplitude factor $T_n$) from the position
$y_n$. Once scattered the particle can go either to left, along
the path $P_{2 n-1}$, or to right, along the path $P_{2 n}$ (see
Fig. 1). To calculate the Green function we have to classify and
to sum up all the scattering trajectories, which always can be
done in a closed form as shown in \cite{marcos1,marcos2}.
Following such references, we find \beq G^N_{j m}(x_f,x_i;k) &=&
\left(\frac{1}{2ik}\right) \frac{{ T}^{(+)}_{j,m+1}\exp[-ik(y_j -
y_{m+1})]}{D_{\mbox{\scriptsize line}}} \left[\exp[-ikx_i] + {
R}^{(-)}_{m,1}
\exp[ikx_i]\exp[-2iky_m] \right] \nonumber\\
&& \times\left[\exp[ikx_f]  + { R}^{(+)}_{N,j+1}
\exp[-ikx_f]\exp[2iky_{j+1}]\right], \label{gline} \eeq where
\beq D_{\mbox{\scriptsize line}} &=& \left[1 - { R}^{(-)}_{m,1} {
R}^{(+)}_{j,m+1}\exp[2ik(y_{m+1}-y_m)] \right] \left[1 - {
R}^{(+)}_{N,j+1} { R}^{(-)}_{j,m+1}\exp[2ik(y_{j+1}-y_j)] \right]
\nonumber\\
&& - { R}^{(-)}_{m,1} { R}^{(+)}_{N,j+1} { T}^{(+)}_{j,m+1} {
T}^{(-)}_{j,m+1}\exp[-2ik(y_j-y_{m+1}+y_m-y_{j+1})].
\label{glinefactor} \eeq The factors $R_{n,l}$ and $T_{n,l}$ are
derived in the Appendix B.

\begin{figure}
\centerline{\psfig{figure=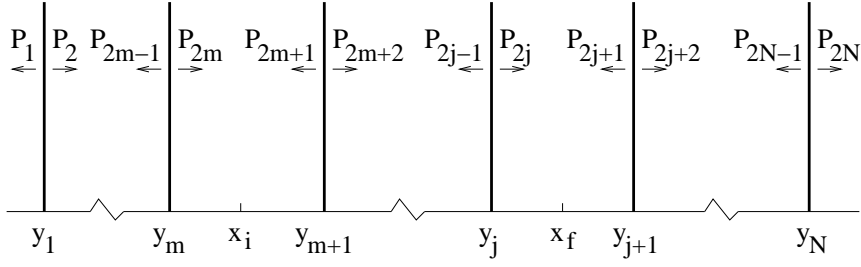,height=3.5cm}}
\caption{$N$ arbitrary point interactions on the line and
located at the arbitrary positions $y$'s.}
\end{figure}

\subsection{Green function for arbitrary number of point
interactions on the half-line}

Recently, an interesting effect, which is related to the so called
atomic mirrors, has been proposed \cite{dodonov}. The idea is to
place a wave-packet initially with a zero mean momentum near a
given barrier potential, e.g., an infinite hard wall or a delta
potential. The wave-packet spreads out and due to energy quantum
fluctuations its mean momentum value increases. This behavior is
associated to an effective "quantum repulsive force", which in
principle can be measured using ultra-cold atoms. The problem is
to reach the necessary extreme conditions in laboratory. Thus, it
would be helpful to enhance the phenomenon, which eventually can
be done by considering more singular potentials. This motivate us
to look at the problem of general point interactions nearby an
infinite hard wall, i.e., $N$ point interactions on the half-line.

The exact Green function for this case is easily obtained from our
previous results. Actually, by setting $y_1=0$ in our general
expression (\ref{gline}) and taking $R_1^{(-)}$ to be -1 or +1, we
have, respectively, Dirichlet or Neumann boundary conditions at
$x=0$. We observe this is simple to consider that the first point
interaction is either $\gamma \delta(x)$ or $\gamma \delta'(x)$,
where the limit $\gamma \rightarrow \infty$ is taken, i.e., the
point interaction becomes an infinity hard wall with different
reflection properties. Just as a simple example, consider the
situation in Fig. 2 (a), an infinite hard wall plus a general
point interaction at $x=y$. Denoting $G_{--}$ for $x_i, x_f < y$
and $G_{+-}$ for $x_i < y < x_f$, we have after straightforward
simplifications (here, $s=1$ for Dirichlet and $s=-1$ for Neumann
boundary conditions)
\begin{eqnarray}
G_{+-}(x_f,x_i;k) &=& \frac{1}{2 i k}
\frac{T^{(+)}}{\left(1+s R^{(+)} \exp[2 i k y]\right)}
\Big\{\exp[i k (x_f - x_i)] - s \exp[i k (x_f + x_i)]\Big\}
\exp[i k y] \nonumber \\
G_{--}(x_f,x_i;k) &=& \frac{1}{2 i k}
\frac{1}{\left(1+s R^{(+)} \exp[2 i k y]\right)}
\Big\{\exp[i k |x_f - x_i|] - s R^{(+)} \exp[2 i k y]
\exp[-i k |x_f - x_i|] \nonumber \\
& & - s \exp[i k (x_f+x_i)] + R^{(+)} \exp[2 i k y]
\exp[-i k (x_f+x_i)] \Big\}.
\end{eqnarray}

\subsection{Green function for arbitrary number of point
interactions inside an infinite well}

A system so simple as a short-range potential placed inside an 1D
infinite well can sometimes present unusual dynamics
\cite{caos-box}. In fact, it has been shown that such system can
exhibit chaotic-like behavior. Another interesting property, which
can be studied in 2D boxes \cite{robinett} as well in systems of
the present type, is the revival time. Initially, a wave-packet is
placed in one side of the box. Then, it evolves spreading over the
whole configuration space it is allowed to fulfill. After a
certain time (the revival time) all the "pieces" of the
wave-packet return to the initial situation, reconstructing the
original state. Such applications may lead one to try to calculate
the Green function for an arbitrary number of point interactions
inside an infinite well.

As before, the exact $G$ for this case can be obtained from
(\ref{gline}).
For so, we just set $y_1=0$, $y_N=L$ and then choose for
$R_1^{(-)}$ and $R_N^{(+)}$ the values -1 or +1. This will lead
to all the four possible combinations of Dirichlet and Neumann boundary
conditions for the walls at $x=0$ and $x=L$.

As an example, consider the potential in Fig. 2 (b), a single
point interaction at $x=y$ inside an infinite well of length $L$.
We restrict the discussion to $G_{--}$ for $x_i, x_f < y$ and
$G_{+-}$ for $x_i < y < x_f$. After lengthy but simple
manipulations, we find the Green functions (again, $s=1$ and
$s=-1$ leads to Dirichlet and Neumann boundary conditions at the
corresponding walls)
\begin{eqnarray}
G_{--}(x_f,x_i;k) &=& \frac{1}{2 i k}
\frac{\left(1 + s_L R^{(-)} \exp[2 i k (L - y)]\right)}{D}
\nonumber \\
& & \times \left\{ \exp[i k |x_f - x_i|] -
s_0 \left( R^{(+)} \exp[2 i k y] -
\frac{s_L T^{(+)} T^{(-)} \exp[2 i k L]}
{1 + s_L R^{(-)} \exp[2 i k(L-y)]} \right)
\exp[-i k |x_f - x_i|] \right. \nonumber \\
& & \left. - s_0 \exp[i k (x_f+x_i)]
+ \left( R^{(+)} \exp[2 i k y] -
\frac{s_L T^{(+)} T^{(-)} \exp[2 i k L]}
{1 + s_L R^{(-)} \exp[2 i k(L-y)]} \right)
\exp[-i k (x_f+x_i)] \right\}, \nonumber \\
G_{+-}(x_f,x_i;k) &=& \frac{1}{2 i k} \frac{T^{(+)}}{D} \Big\{
\big(\exp[- i k x_i] - s_0 \exp[i k x_i]\big) \big(\exp[i k x_f] -
s_L \exp[2 i k L] \exp[-i k x_f]\big) \Big\},
\end{eqnarray}
with
\begin{equation}
D = (1 + s_0 R^{(+)} \exp[2 i k y])
(1 + s_L R^{(-)} \exp[2 i k (L-y)]) -
s_0 s_L T^{(+)} T^{(-)} \exp[2 i k L].
\end{equation}

The poles of $G$ give the system eigenvalues. In our case, they
come from $D = 0$.

\begin{figure}
\centerline{\psfig{figure=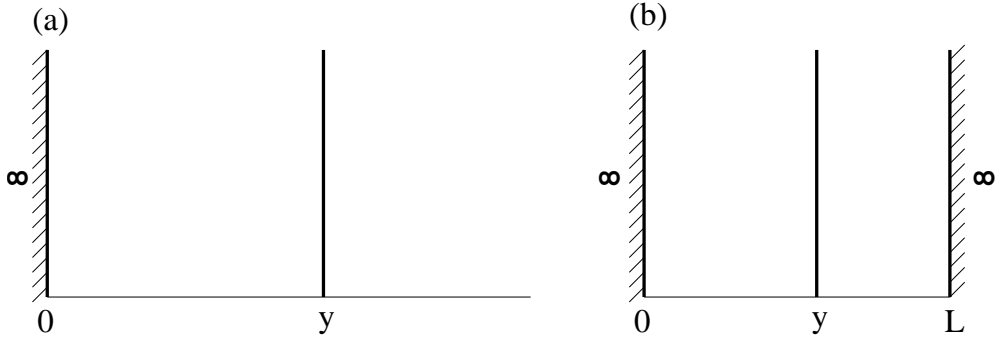,height=4.5cm}}
\caption{A general point interaction at $y$; (a) on the
half-line; and (b) within in infinite well. In both cases
we may have Dirichlet or Neumann boundary conditions at the
walls.}
\end{figure}

\subsection{Green function for periodic boundary conditions}

Here we consider periodic boundary conditions for the wave
function, or $\psi(-L/2) = \psi(L/2)$ and $\psi'(-L/2) =
\psi'(L/2)$. So, the Green function satisfy exactly these same
relations in both $x_i$ and $x_f$. For simplicity, let we start
with one point potential at $x=0$, see Fig. 3 (a). We need to
consider all the scattering paths starting from $x_i$ and arriving
at $x_f$. We can think of the points $-L/2$ and $L/2$ as being
connected, so we have trajectories on a circle. The paths are then
given by arbitrary number of rounds clockwise and anti-clockwise
on the circle. Each time the particle hits the point interaction
at $x=0$, it is either reflected from or transmitted through the
potential (with the amplitude corresponding to that trajectory
getting the factor $R$ or $T$, respectively). The paths are
continuous and no extra factor to construct $W_{\mbox{\scriptsize
s.p.}}$ is gained when the particle cross $-L/2$ to $L/2$ or $L/2$
to $-L/2$. By classifying and summing up all the scattering paths
we finally find the exact Green function as
\begin{eqnarray}
G(x_f,x_i;k) &=& \frac{1}{2ik} \frac{1}{D_{\mbox{\scriptsize circle}}}
\Big\{
\left[ T^{(+)} + \left( R^{(+)} R^{(-)} - T^{(+)} T^{(-)} \right)
\exp[i k L] \right] \exp[i k (x_f - x_i)] \nonumber \\
& & +
\left[1 - T^{(+)} \exp[i k L] \right] \exp[i k (L - (x_f - x_i))]
\nonumber \\
& &
+ R^{(+)} \exp[i k (L - (x_f + x_i))] +
R^{(-)} \exp[i k (L + (x_f + x_i))] \Big\},
\end{eqnarray}
with
\be
D_{\mbox{\scriptsize circle}} = \left(1 - T^{(+)} \exp[ikL]\right)
    \left(1 - T^{(-)} \exp[ikL]\right) -
R^{(+)} R^{(-)} \exp[2 i k L].
\ee
The zeros of $D_{\mbox{\scriptsize circle}}$ are the energy eigenvalues.

\begin{figure}
\centerline{\psfig{figure=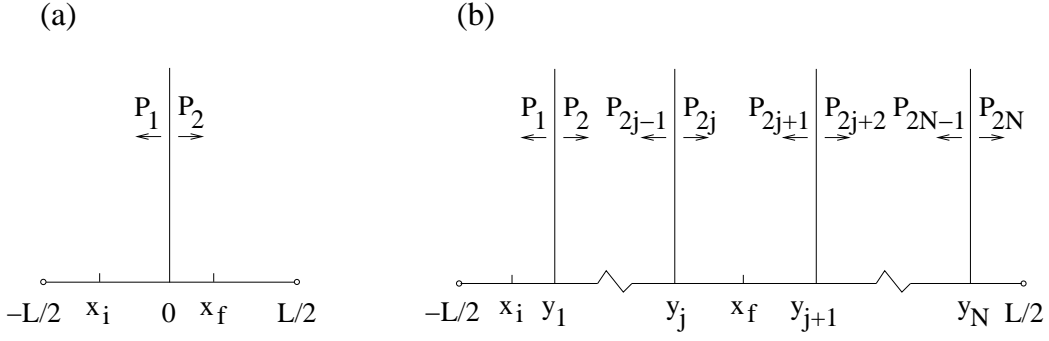,height=4.5cm}}
\caption{Periodic boundary conditions, i.e., the point $x=-L/2$ is
equivalent to the point $x=L/2$. (a) A single general point interaction
at $x=0$. (b) $N$ general point interactions located at
arbitrary positions $y$'s.}
\end{figure}

The same reasoning to construct the scattering paths can be used
to the more general case shown in figure 3 (b). The final result
is \beq G(x_f,x_i;k) &=& \frac{1}{D_{\mbox{\scriptsize pbc}}}
\left\{ \left[{ T}^{(+)}_{j,1}\exp{[-ik(y_j-y_1)]} + {
T}^{(-)}_{N,j+1}{ K}_{j,1}\exp{[-ik(y_N-y_{j+1})]} \exp[ikL]
\right] \exp{[ik(x_f-x_i)]}
\right. \nonumber \\
& & + \left[{ R}^{(-)}_{N,j+1}{
T}^{(+)}_{j,1}\exp[ik(L-2y_N-y_j+y_1)] + { R}^{(-)}_{j,1}{
T}^{(-)}_{N,j+1}\exp[-ik(2y_j+y_N-y_{j+1})]\right]
\exp[ikL]\nonumber\\ &&\times \exp{[ik(x_f+x_i)]}
\nonumber \\
& & +\left[ { R}^{(+)}_{j,1}{
T}^{(-)}_{N,j+1}\exp[ik(L+2y_1-y_N+y_{j+1})] + { R}^{(+)}_{N,j+1}{
T}^{(+)}_{j,1}\exp[ik(2y_{j+1}-y_j+y_1)] \right]
\exp{[-ik(x_f+x_i)]}
\nonumber\\
&& +  \left[ { T}^{(-)}_{N,j+1}\exp[-ik(y_N-y_{j+1})] + {
T}^{(+)}_{j,1} { K}_{N,j+1}\exp[-ik(y_j-y_1)] \exp[ikL]\right]
\exp[ikL] \nonumber\\
&&\times \left.\exp{[-ik(x_f-x_i)]} \right\},
\eeq
where ${K}_{b,a} = { R}^{(+)}_{b,a}{ R}^{(-)}_{b,a}
\exp[2ik(y_a-y_b)] - {
T}^{(+)}_{b,a}{ T}^{(-)}_{b,a}\exp[-2ik(y_b-y_a)]$ and \beq
D_{\mbox{\scriptsize pbc}} &=& -1 +{ R}^{(+)}_{j,1}{
R}^{(-)}_{N,j+1} \exp[2ik(L-y_N+y_j)] + { R}^{(-)}_{j,1}{
R}^{(+)}_{N,j+1}\exp[2ik(y_{j+1}-y_j)]
\nonumber\\
&&- \left({ T}^{(-)}_{j,1}{ T}^{(-)}_{N,j+1} +{ T}^{(+)}_{j,1}{
T}^{(+)}_{N,j+1}\right) \exp[-ik(y_N-y_{j+1}+y_j-y_1)]  - {
K}_{j,1}{ K}_{N,j+1} \exp[2ikL]. \eeq For some of the above
quantities see the Appendix B.

\section{Remarks and Conclusion}

Here we have presented a new way to calculate the Green functions
for generalized point interactions in 1D. Our approach strikes out
in a different direction than most theoretical treatments. The
method is based mostly on physical grounds. First, we show that
one can define a general point interaction through its scattering
properties, i.e., the reflection and transmission amplitudes.
Then, we discuss how to construct the exact Green function from
such coefficients. It helps to keep track of the relevant physical
quantities what may not happen in some other methods due to their
subtle renormalization procedures.

We do not invoke any kind of regularization, such as
renormalization, series expansion or self-adjoint extension. The
advantage in not using specific mathematical prescription is
that in general such techniques may not have a clear physical
meaning (differently from some other contexts, where there are
solid principles as a guide for regularization, like in the case
of Mandelstam-Leibbrandt prescription for light-cone gauge in
field theory) or are not unique. In fact, one can find in the
literature the Green function for a same general point interaction
based on calculations from either a single $\delta'$ potential
\cite{park2} or a combination of $\delta$ and $\delta'$ potentials
\cite{albeverio}. This apparent contradictory result is just due
to the fact that the authors use different prescriptions to
regularize their Hamiltonians.

The second advantage of this more pedestrian method is that we can
easily extend our calculations to more general cases. Indeed, we
have derived the Green functions for $N$ generalized point
interactions for several situations; (i) on a line; (ii)
restricted to the half-line; and (iii) within in an infinite box.
For the later two cases we can impose different boundary
conditions at the border walls. Also, (iv) we have obtained $G$
for periodic boundary conditions (the circle-like case). As far as
we know, explicitly expressions for $G$ for all these systems were
not known in the literature and our aim was to fulfill this gap.

A way to construct generalized point interactions has been
recently proposed in a series of papers
\cite{shigehara-pla,fb-duality}. The idea is to consider different
usual delta functions, all separated by a distance $y_0$, with
appropriate values for their strengths. In the limit of $y_0
\rightarrow 0$, one obtains the desired potential. Although the
exact limit cannot be taken in practice, if we can approximate
each delta  by some short range potential and then have $y_0$
finite but small, we may in principle obtain a physical
realization of a general point interaction. This open a great
possibility of experiments in order to test fundamental and
interesting phenomena in quantum mechanics as applications for the
systems discussed in this work. In particular, of great interest
would be the calculation of the time evolution of wave-packets,
 $\Psi(x_f,t) = i/(2 \pi)
\int \int dx_i dE \exp[-i E t] G(x_f,x_i;E) \Psi(x_f,0)$, or of
the density of states
$\rho(E) = -\frac{1}{\pi}\Im \int dx\; G(x,x;E)$.
Due to the form of our Green function, the integration in the
position can be solved analytical for $\rho$. For $\Psi$, it
can also be done for simple initial wave-packets like Gaussians.
The integral on the energy, by its turn, can be easily carried out
by using numerical methods such as the FFT.

Regarding extensions of the present results we comment the
following. In Ref. \cite{duality}, both the local description of
point interaction (which we used in this work) and a global one
based on $U(2)$ group (which contains the former) were discussed.
In this $U(2)$ context, one can also obtain the reflection and
transmission coefficients. So, our result are valid in the global
description of point interactions. We emphasize that the key point
to obtain all the Green functions calculated here is to know the
$R$'s and $T$'s coefficients of the potentials. In fact, our
results apply for every one-dimensional $V(x)$ which satisfies the
assumptions given in \cite{marcos1,marcos2}. Thus, a ``mixing'' 1D
lattice including both general point interactions and usual
potentials (which decay at least exponentially) can be calculated
from our method.

As a final remark we mention that it would be interesting to extend
the present approach to higher dimensions. In a recent work
\cite{luz-heller}, it has been discussed how to calculate the wave
function and also the Green function for boundary walls of
arbitrary shapes and with very general boundary conditions. The
method is based on the calculation of a $T$ matrix, which plays
a similar role than the 1D quantum amplitudes. By taking appropriate
limits, we could construct point like interactions in 2 and 3
dimensions for which the Green function is already regularized.
Such work is in progress and will be reported in the due course.

\begin{acknowledgments}
Schmidt and Luz gratefully acknowledge CNPq for research fellowships.
\end{acknowledgments}

\appendix

\section{Special Cases}

In order to exemplify our general result for a single generalized
point interaction on the line,  eq. (\ref{green-aberto}),
we show how to obtain from it the well-known Green functions for
delta \cite{blinder} and delta-prime
\cite{park2,dirac,grosche1} potentials.

The delta potential is a particular case of
(\ref{condicao-geral}), where $a=d=\omega=1, b=0, c=\gamma$.
Substituting these parameters into eqs. (\ref{rt}) and
(\ref{green-aberto}), one obtain

\be \label{green-aberto-delta} G_{\pm \mp} = \frac{\exp[ik |x_f -
x_i|]}{2ik-\gamma} , \qquad\qquad G_{\pm \pm} = \frac{1}{2ik}
\left[ \exp[ik |x_f - x_i|] +
\left(\frac{\gamma}{2ik-\gamma}\right)  \exp[i k( |x_f| +
|x_i|)]\right] . \ee
Now, recalling the meaning of the subscripts for $G$ (see Sec. II)
in the above eq., it is not difficult to realize that we can write
(\ref{green-aberto-delta}) in the following single formula
\be G = \frac{1}{2ik}
\left[ \exp[ik |x_f - x_i|] +
\left(\frac{\gamma}{2ik-\gamma}\right)  \exp[i k( |x_f| +
|x_i|)]\right], \ee
that agrees with Eq. (17) of \cite{blinder}
if we identify in the coefficients which multiply the exponentials:
$\gamma \leftrightarrow -Z$ and $2 i k \leftrightarrow i k$ (this last
relation is due to the fact that in \cite{blinder} the author uses
$m=1$ instead of $m=1/2$ as in our case).

The delta prime potential is defined by
$a=d=\omega=1, b=\gamma, c=0$, so one get the Green functions

\be \label{green-aberto-delta-prime} G_{\pm \mp} = \frac{\exp[ik
|x_f - x_i|]}{(2i+\gamma k)k}, \qquad\qquad G_{\pm \pm} =
\frac{1}{2ik} \left[ \exp[ik |x_f - x_i|] + \left(\frac{\gamma
k}{2i+\gamma k}\right) \exp[i k( |x_f| + |x_i|)]\right]. \ee
Again, it is easy to show that the above expressions can be
summarized as (sign(.) is the signal function)
\be
G = \frac{1}{2ik}
\left[ \exp[ik |x_f - x_i|] + \left(\frac{\gamma
k}{2i+\gamma k}\right) \mbox{sign}(x_f) \, \mbox{sign}(x_i)
\exp[i k( |x_f| + |x_i|)]\right].
\label{gdeltap}
\ee
Equation (\ref{gdeltap}) agrees with Eq. (12) of Grosche
in \cite{dirac}
if we identify: $\gamma \leftrightarrow -\beta$,
$-ik \leftrightarrow \sqrt{-E}$ (because in such reference
it is used a Wick rotation) and $G \leftrightarrow -G$ (due
to an opposite sign used in the definition of the Green function).

\section{Recurrence Formulas}

Here we show how to obtain some coefficients used in the exact
expressions for the Green functions by means of recurrence formulas.
The idea is to face a series of $n-l+1$ potentials located at
$y_l, y_{l+1}, \ldots, y_{n-1}, y_n$ as a single block
and then to associate to it the amplitudes $R_{n,l}$ and $T_{n,l}$
(for a detailed discussion see Ref. \cite{marcos1}).

Assume firstly a potential composed by two point interactions,
placed at $y_l$ and $y_{l+1}$, and let $x_i, x_f < y_l < y_{l+1}$.
By using the same approach developed throughout this paper, we
find for the Green function
\be G_{--} =
\exp[ik |x_f-x_i|] +
R^{(+)}_{l} \exp[-ik(x_f+x_i- 2 y_l)] +
\frac{R^{(+)}_{l+1}T^{(+)}_{l}T^{(-)}_{l}
\exp{[-ik(x_f+x_i-2 y_l)]}}
{1-R^{(-)}_{l} R^{(+)}_{l+1}\exp[2ik(y_{l+1}-y_l)]}.
\ee
We define a reflection coefficient for this block (made by
two potentials) as
\be
\label{r2-recor}
R^{(+)}_{l+1,l} = R^{(+)}_{l} +
\frac{R^{(+)}_{l+1} T^{(+)}_{l} T^{(-)}_{l} \exp[2ik(y_{l+1}-y_l)]}
{1 - R^{(-)}_{l} R^{(+)}_{l+1}\exp{[2ik(y_{l+1}-y_l)]}}.
\ee
In analogy, we can define $R^{(-)}_{l+1,l}$ for this same block
by calculating $G$ for $x_i, x_f > y_{l+1} > y_l$.

Now, consider $x_i < y_{l} < y_{l+1} <x_f$, we have
\be
G_{+-} = \frac{T^{(+)}_{l}T^{(+)}_{l+1} \exp[ik(y_{l+1}-y_l)]}
{1 - R^{(-)}_{l} R^{(+)}_{l+1} \exp{[2ik(y_{l+1}-y_l)]}}
\exp[ik(x_f-x_i - (y_{l+1}-y_l))],
\ee
thus, we again can define
\be
T^{(+)}_{l+1,l} = \frac{T^{(+)}_{l}T^{(+)}_{l+1}
\exp{[ik(y_{l+1}-y_l)]}}
{1-R^{(-)}_{l} R^{(+)}_{l+1} \exp{[2ik(y_{l+1}-y_l)]}}.
\ee
Similarly, we derive $T_{l+1,l}^{(-)}$ by calculating
$G$ for $x_i > y_{l+1} > y_l > x_f$.

In order to get recurrence formulas consider a third potential
located at $y_{l+2}$
(recall that our two point interactions block has its ending points
at $y_l$ and $y_{l+1}$).
Let $x_i, x_f < y_l < y_{l+1} < y_{l+2}$ and the $R_{l+1,1}$ and
$T_{l+1,l}$
of the two potentials block as obtained before.
By a direct inspection in (\ref{r2-recor}), we readly can infer the
reflection coefficient $R^{(+)}_{l+2,l}$
(for the new block formed by the three potentials) as
\be
\label{r3-recor}
R^{(+)}_{l+2,l} =
R^{(+)}_{l+1,l} +
\frac{T^{(+)}_{l+1,l} T^{(-)}_{l+1,l}
R^{(+)}_{l+2} \exp[2ik(y_{l+2}-y_{l+1})]}
{1 - R^{(-)}_{l+1,l} R^{(+)}_{l+2} \exp[2ik (y_{l+2}-y_{l+1})]},
\ee
the generalization is then straightforward
\be
R^{(+)}_{n,l} = R^{(+)}_{n-1,l} +
\frac{T^{(+)}_{n-1,l} T^{(-)}_{n-1,l} R^{(+)}_{n}
\exp[2ik(y_n-y_{n-1})]}
{1 - R^{(-)}_{n-1,l} R^{(+)}_{n} \exp[2ik (y_n-y_{n-1})]}.
\ee
For the reflection coefficient $R^{(-)}_{n-1,l}$ one finds
\be
R^{(-)}_{n,l} = R^{(-)}_{n} +
\frac{T^{(+)}_{n} T^{(-)}_{n} R^{(-)}_{n-1,l}
\exp[2ik (y_n - y_{n-1})]}
{1- R^{(-)}_{n-1,l} R^{(+)}_n \exp[2ik(y_n-y_{n-1})]}.
\ee

Transmission coefficients can be written in terms of recurrence
relations as well. The final results are
\be
T^{(+)}_{n,l} =
\frac{T^{(+)}_{n-1,l} T^{(+)}_{n} \exp[i k (y_n-y_{n-1})]}
{1- R^{(-)}_{n-1,l} R^{(+)}_{n} \exp[2ik (y_n - y_{n-1})]},
\ee
and
\be
T^{(-)}_{n,l} =
\frac{T^{(-)}_{n-1,l} T^{(-)}_{n} \exp[i k (y_n-y_{n-1})]}
{1- R^{(-)}_{n-1,l} R^{(+)}_{n} \exp[2ik (y_n-y_{n-1})]}.
\ee

\end{document}